\documentstyle[epsfig]{article}

\input ibvs2.sty

\begin{document}

\IBVShead{}{}

\IBVStitle{Planetary Transits of TRES-1}

\IBVSauth{Price, A.$^{1}$; Bissinger, R.$^{1}$; Laughlin, G.$^{2}$; Gary, B.$^{1}$; Vanmunster, T.$^{3}$;
Henden, A.$^{4}$; Starkey, D.$^{1}$; Kaiser, D.$^{1}$; Holtzman, J.$^{5}$; Marschall, L.$^{6}$; Michalik, T.$^{7}$; Wellington, T.$^{7}$;
 Paakkonen, P.$^{8}$; Kereszty, Z.S.$^{1}$; Durkee, R.$^{9}$; Richardson, K.$^{10}$; Leadbeater, R.$^{1}$; 
Castellano, T.$^{11}$
   }

\IBVSinst{American Association of Variable Star Observers (AAVSO), \\
\indent Clinton B. Ford Astronomical Data \& Research Center, 25 Birch St., Cambridge, MA 02138, USA; \\
\indent email: aavso@aavso.org}

\IBVSinst{UCO/Lick Observatory, Department of Astronomy and Astrophysics \\
\indent University of California, Santa Cruz, CA 95064 USA; \\
\indent email: laugh@ucolick.org}

\IBVSinst{Center for Backyard Astrophysics - Belgium, Walhostraat 1A, B-3401 Landen, Belgium\\
\indent e-mail : Tonny.Vanmunster@cbabelgium.com}

\IBVSinst{Universities Space Research Association/U. S. Naval Observatory, Flagstaff, AZ 86001 USA;\\
\indent email: aah@nofs.navy.mil}

\IBVSinst{New Mexico State University, Box 30001 / Department 4500, 1320 Frenger St. Las Cruces, NM 88003 USA  \\
\indent email: holtz@nmsu.edu}

\IBVSinst{Department of Physics, Gettysburg College, Gettysburg, PA 17325 USA \\
\indent email:  marschal@gettysburg.edu}

\IBVSinst{Randolph-Macon Woman's College,  2500 Rivermont Avenue, Lynchburg, VA 24503 USA\\
\indent email: tmichalik@rmwc.edu, tawellington@rmwc.edu}

\IBVSinst{Amateur Astronomers Association Seulaset, Finland}

\IBVSinst{Shed of Science Observatory, Minneapolis, MN 55410 USA\\
\indent email: shedofscience@earthlink.net }

\IBVSinst{San Diego Astronomy Association, San Diego, CA  USA \\
\indent e-mail: kentr@san.rr.com web: www.sdaa.org}

\IBVSinst{NASA Ames Research Center,  Moffett Field, CA 94035 USA \\
\indent email: tcastellano@mail.arc.nasa.gov }

\SIMBADobj{GSC02652-01324}

\IBVStyp{}
\IBVSkey{}

\IBVSabs{}
\IBVSabs{}
\IBVSabs{}
\IBVSabs{}
\IBVSabs{}

\begintext

 The Trans-Atlantic Exoplanet Survey (TrES) has announced the discovery of a planet, 
having 0.75 +/-0.07 Jupiter masses, transiting the star GSC02652-01324 (2MASS J19040985$+$3637574), a V=11.79 magnitude, K0 V type star
at $\alpha$: 19:04:09.8 $\delta$: +36:37:57 (J2000) (Alonso et al. 2004).
Transitsearch.org (Seagroves 2003), in collaboration
with the AAVSO, compiled observations from professional and amateur astronomers of predicted
transits in the summer and autumn of 2004. Analysis of the photometry suggests possible post-eclipse 
brightening episodes after the egress portion of the TRES-1b light curve. 

 The AAVSO compiled 10,560 CCD observations covering seven complete TRES-1b predicted transit windows, 
three windows of partial coverage and coverage of baseline non-transit periods. 
(Figure 1; Table 1)

 The amateur observations
were done individually using a variety of equipment, comparison stars and reduction software. 
Different filters were also used. For the plotting of the phase diagram (Figure 2) the observations were 
offset to a common zero point. Uncertainties for the individual observations are available along with the entire raw data set upon
request to the AAVSO\footnote{Data can be downloaded from http://www.aavso.org/data/download/}. 

 Additional photometry were provided by three observatories. The U.S. Naval Observatory Flagstaff Station 1.0m telescope 
observed {\it B} using with a large set of Landolt standards (Landolt 1992) having a range of color and airmass. 
IRAF was used for all reduction.  Observations were also provided by the Gettysburg College Observatory with 
a Photometrics CH-350 camera (SiTE 003B back illuminated 1024K chip) on a 16" f/11 reflector, with standard
Bessell filters (Bessell 1995). IRAF was used for flat-fielding, dark correction, 
bias subtraction, and MIRA for the photometry. In addition, The New Mexico State University 1m at Apache Point 
Observatory observed using an Apogee AP7p camera with a 512x512 thinned backside illuminated SITe chip.

\IBVSfig{20cm}{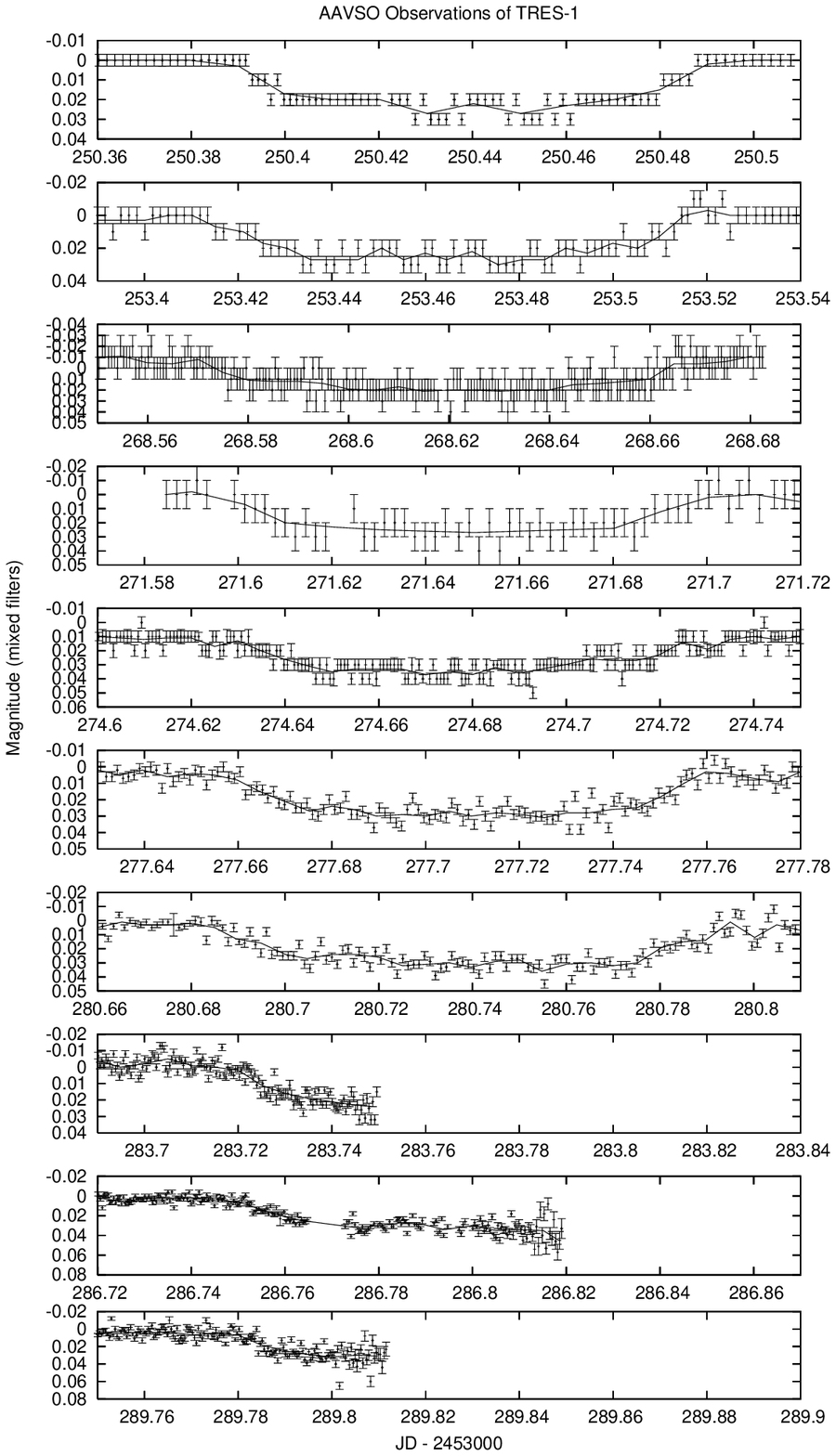}{Light curves of ten transits. Average line is 0.005d and egress brightenings noted with verticle line. 
For readability, lower precision data was omitted when higher precision 
coincident data was available. All measurements and individual uncertainty values are available from www.aavso.org. }


\begin{table}[h]\centering
\begin{tabular}{|c|c|c|c|c|} \hline 
{\it Transit Date (JD UT)} & {\it Observer(s) (Filter)} \\ \hline
 2453250  & Vanmunster{\it (V)} \\ \hline
 2453253  & Paakkonen{\it (V)}, Starkey{\it (V)}, Vanmunster (Unfiltered) \\ \hline
 2453256  & Leadbeater (Unfiltered) \\ \hline
 2453268  & Kaiser{\it (V)}, Marschall{\it (R)}, Michalik{\it (V)} \\ \hline
 2453271  & Durkee{\it (V)}, Gary{\it (V)}, Starkey{\it (V)} \\ \hline
 2453274  & Henden{\it (B)} \\ \hline
 2453277  & Bissinger (Unfiltered) \\ \hline
 2453280  & Bissinger (Unfiltered), Gary{\it (V)} \\ \hline
 2453283  & Bissinger (Unfiltered)\\ \hline
 2453286  & Bissinger (Unfiltered), Gary{\it (V)}, Holtzman{\it (V)} \\ \hline
 2453289  & Bissinger (Unfiltered)\\ \hline
\end{tabular}
\caption{Observed TrES-1b Transit Windows}
\end{table}

Visual inspection of the light curves reveals the presense of slight humps at the egress points of some transits (Figure 3). 
These humps have amplitudes of around 0.005 magnitude and occur around 90 minutes after the transit midpoint.
While most individual observations had uncertainties greater than 0.005 magnitudes, when combined the uncertainty 
can be reduced by a factor inversely proportional to the square root of the number of stacked observations. To determine 
the statistical significance of the humps a  
boot strap Monte Carlo simulation was applied to the data. Given a suitably large sample size, the bootstrap method 
redistributes the existing data randomly, generating a synthetic data set. If done numerous times the 
procedure allows uncertainties in the data to be quantified without any assumption of a 
Gaussian distribution. A model for the egress portion of the transit was developed:

\begin{equation}
m=0.1cos(0.071t+0.5)e^{-0.040t}+1 
\end{equation}
 
where {\it m} is the amplitude in magnitudes and {\it t} is the time from the transit midpoint in minutes. 
Bootstrap runs were made on the datasets and post curve fit models using 2,000 trials. The results
confirm that the visually detected hump exists to a statistically significant degree. Bissinger's dataset 
was tested twice for verification and emerged with a mean of 1.01026 (95\% confidence level) and 1.01025 (95\% confidence level), indicating a high level 
of reproducibility. High-precision photometry during the 2005 observing season is needed to confirm the brightening episodes.

The AAVSO acknowledges the Curry Foundation for funding of the CCD equipment used in observations 
by Kereszty.

\IBVSfig{6cm}{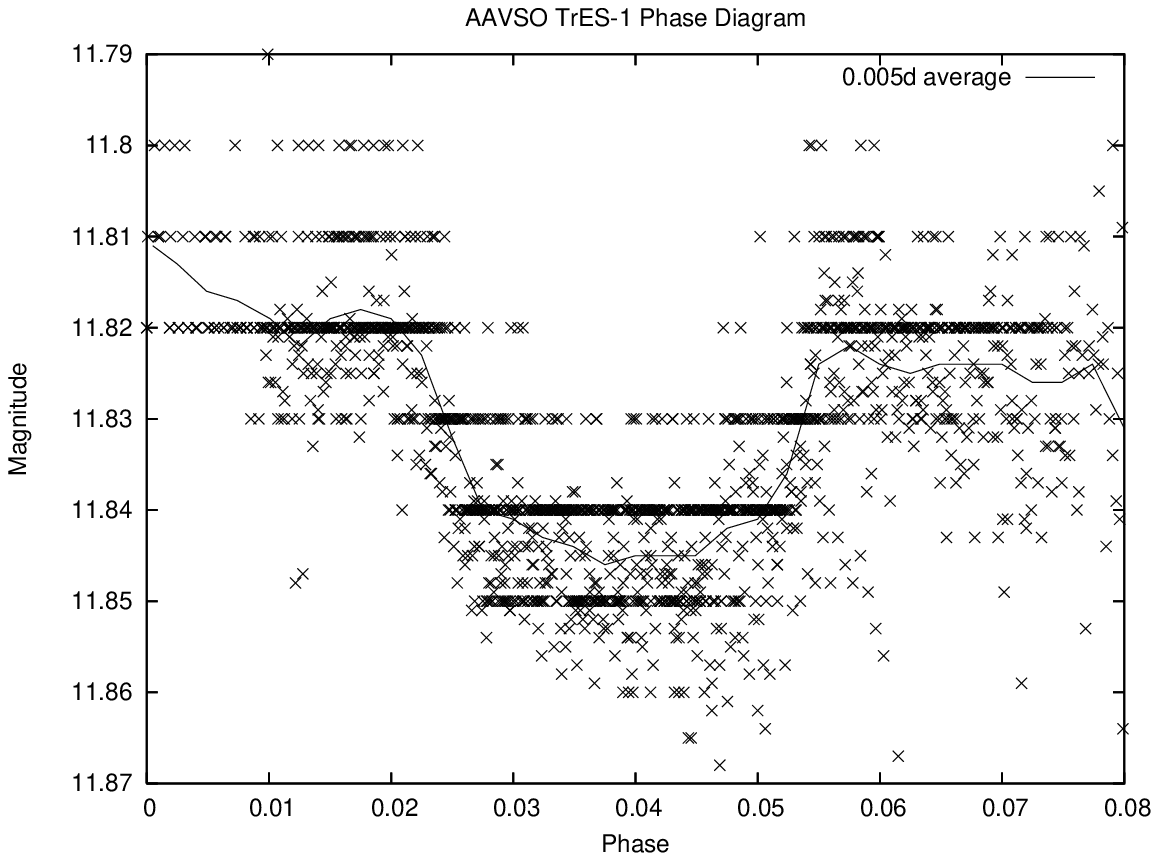}{Phase diagram with a 3.030065d period beginning at epoch JD 2453250.3204. 
Observations were a mix of filtered and unfiltered data which was offset to a common zeropoint. Banding is caused 
by observers who report 0.01 magnitude accuracy.}

\IBVSfig{6cm}{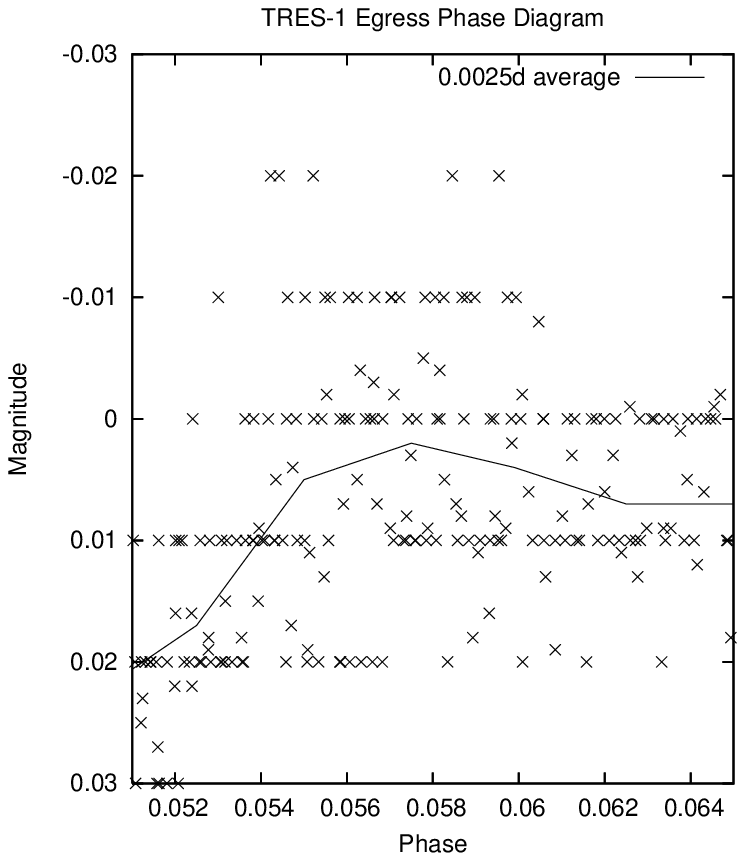}{Detail of the phase diagram in Figure 2 to show brightening during the egress.}

\references

Alonso R., Brown, T.M., Torres G. et al., 2004, http://arxiv.org/abs/astro-ph/0408421 \BIBCODE{}

Seagroves, S., Harker, J., Laughlin, G., Lacy, J., Castellano, T. 2003, {\it Publ. Astron. Soc. Pac.}, {\bf 115}, 1355. \BIBCODE{2003PASP..115.1355S}

Landolt, A. 1992, {\it Astron. Jour.}, {\bf 104}, 340  \BIBCODE{1992AJ....104..340L}

Bessell, M.S. 1995, {\it CCD Astron.}, {\bf 2}, 20. \BIBCODE{1995CCDA....2...20B}

\endreferences

\end{document}